# Open Collaboration for Innovation: Principles and Performance

SHEEN S. LEVINE, Columbia University
MICHAEL J. PRIETULA, Emory University

1. WHAT IS OPEN COLLABORATION?

Open source software is booming. Once the domain of hobbyists and hackers, it has gained acceptance with consumers, corporations, and governments. Some exemplars of open source, such as the *Linux* and *Android* operating systems, are now commonplace, operating millions of devices. Together with other products of open source software, they have been creating billions of dollars in economic value (European Commission 2006).

Yet the same patterns of collaboration, innovation, and production can now be found beyond software (Baldwin and Von Hippel 2011; Benkler 2006; von Hippel 2005b). For example, people collaborate, sometimes with complete strangers, in user-to-user forums (Lakhani and von Hippel 2003), mailing lists (Jarvenpaa and Majchrzak 2008) and online communities (Faraj and Johnson 2010). Some share openly (and occasionally illegally) digital media: music, movies, TV programs, software (Levine 2001). People also share processing power and internet bandwidth, enabling free services such as *Skype* (Benkler 2006, pp. 83-7). In the physical world, off the internet, people give, receive, and share tools and appliances (Goodman 2010; Nelson et al. 2007; Willer et al. 2012), even host strangers overnight (Lauterbach et al. 2009; Perlroth 2011), all without payments or barter. Such ventures exemplify what we call *open collaboration* (OC), a shorthand inspired by Baldwin and von Hippel (2011). Here we define its principles and explore its performance.

Firms have been affected by open collaboration, some negatively, others positively. The free encyclopedia *Wikipedia*, a prime example of such collaboration, has come to match the quality of *Encyclopædia Britannica* (Giles 2005), which, after 244 years in circulation, has ceased printing. Other firms have been thriving by facilitating open collaborations, hosting forums and communities. This is how firms such as *Amazon*, an internet retailer, and *TripAdvisor*, a review site for hotels and restaurants, established a flow of "user generated content": reviews, advice, photos, and video clips. Fellow users may benefit from such information, and the firms economize on wage-free, royalty-free content (Chevalier and Mayzlin 2006; Mudambi and Schuff 2010). The effect of open collaboration may be nascent, but it has already required established firms to tweak their strategy, operations, and marketing (Chen and Xie 2008; Scott and Orlikowski 2012). It also has an impact outside the commercial world.

Open collaboration fits well with the scientific ethos (David 1998), and — not surprisingly — it has been benefiting scientific endeavors. For instance, thousands of volunteers, each contributing just a fraction of a solution, have been discovering and solving problems too immense for traditional organizations (Benkler 2004; Partha and David 1994). Contributors classify celestial objects in *Galaxy Zoo*, decipher planetary images in the *Mars Public Mapping Project* (Benkler 2006, p. 69), and labor over terrestrial maps (Helft 2007). Similar patterns that are labeled with the adjectival "open" have appeared in medicine (Ortí et al. 2009; Rai 2005), engineering ("open design") and biotechnology (Henkel and Maurer 2007, 2009). In scientific publication, the Open Science movement aims to disperse authority and expand collaboration (Lin 2012).

Scholars have been attracted to these novel patterns of innovation and production. Open source software, a harbinger of open collaboration, was the topic of several edited volumes (e.g., Feller et al. 2005; West and Gallagher 2006) and special issues of journals (von Krogh and von Hippel 2003, 2006). The use of "open source" as a scholarly term has been growing dramatically, from just 32 appearances in 1999 to 687 times a decade later (see in Electronic Companion). We build on these efforts. Here we extend and generalize what prior research called *Open Collaborative Innovation Projects* (Baldwin and Von Hippel 2011), *Peer Production* (Benkler 2002), *Community Based Innovation System* (Franke and Shah 2003), *Wikinomics* and *Mass Collaboration* (Tapscott and Williams 2006), as well as instances of *Collaborative communities* (Adler et al. 2008), *Transaction-Free Zones* (Baldwin 2008), *Crowdsourcing* (Afuah and Tucci 2012), *Collaborative Consumption* (Goodman 2010), *Electronic Networks of Practice* or *Online Communities* (Faraj et al. 2011; Kollock 1999; Wasko and Faraj 2005), and *Open Innovation* (West and Gallagher 2006).

Such open collaborations have drawn scholarly interest because of their social and economic impact. However, what affects their performance – even why they are viable – remains a puzzle. We begin by identifying some defining principles: a system of innovation or production that relies on goal-oriented yet loosely coordinated participants, who interact to create a product (or service) of economic value, which is made available to contributors and non-contributors alike.

## 2. WHAT AFFECTS THE PERFORMANCE OF OPEN COLLABORATION?

### 2.1 The human tendency to cooperate

Next, we examine performance. We investigate several elements that affect the performance of open collaboration: One element is cooperative behavior of contributors, who willingly share their work (or property) with non-contributors. Performance can benefit with people benefit others at cost to themselves, even without guarantee of reciprocity. But it is also inherently risky — contributors can be overwhelmed by free riders, as in the classic Tragedy of the Commons (Hardin 1968; Olson 1965). Thus, we draw on recent evidence on human cooperation (Ishii and Kurzban 2008; Kurzban and Houser 2005) to answer a fundamental question: why people share the fruits of cooperation with non-contributors.

Early theoreticians presumed that, without safeguards, cooperation would be displaced by free riding. The prevalent presumption was that "every agent is actuated only by self-interest" (Edgeworth 1881, p. 16). When OC emerged, scholars sought to explain it as self-interest. Participants must have some direct and immediate benefit from doing so, the thinking went, or else they never would have contributed.

However, self-interest is common but not omnipresent. Scholars, including economists, have long questioned whether humans are truly defined by narrow self-interest (e.g., Dawes and Thaler 1988; Sen 1977). In recent years, a more nuanced picture has emerged. Aiming to quantify the frequency and distribution of cooperation and reciprocity, Kurzban and Houser (2005) used an elaborate experimental design to study cooperation between- and within-individuals. They found that people are consistent in the extent of their cooperation in different situations. People's behavioral types are so stable as to allow accurate prediction: "a group's cooperative outcomes can be remarkably well predicted if one knows its type composition" (Kurzban and Houser 2005, p. 1803). The general human population has been estimated to consist of three cooperative types: *Cooperators* (13% of the general population), *Reciprocators* (63%), and *Free Riders* (20%). (The remaining 4% are too inconsistent to be categorized).

### 2.1 Need heterogeneity of participants

Cooperativeness may be the life-giving element of OC, but it is not the only element affecting performance. We add to cooperation, a characteristic of interaction, other two elements taken from the economics and innovation literature. One is a characteristic of the participants: Need Heterogeneity, which is the extent to which they have heterogeneous (diverse) needs. OC participants are diverse, empirical accounts show, and differ in the resources they seek and goods they produce (e.g., Anthony et al. 2009, p. 283; Benkler 2004, p. 1110; Jeppesen and Lakhani 2010; Madey et al. 2004). Researchers have documented the pattern across a variety of products and services: people's needs are often homogenous (see review in Franke et al. 2009; von Hippel 2005a, pp. 33-43). How diversity affects OC performance was of considerable discussion. Some academics (Bonaccorsi and Rossi 2003, p. 1244) and practitioners (Ghosh 1998) argued that diversity supports OC. Higher heterogeneity of needs, they proposed, leads to better performance (von Hippel 2005a, pp. 33-43). It appears plausible: as Platt (1973) pointed out, the Tragedy of The Commons occurs because too many individuals seek the same good; the tragic outcome is caused by low need heterogeneity. When people seek a diversity of goods, the commons can thrive. But at least one study suggested the opposite (Baldwin et al. 2006). Similar needs, it was argued, allow users to solve a problem only once, and share the solution, avoiding duplicate efforts. In our experiment, we consider main and interaction effects separately, so we can show how these differing propositions about diversity and performance can be settled.

### 2.2 Rivalry of goods and resources

Another a characteristic of the goods: Rivalry (aka subtractability, Hess and Ostrom 2006), which is the extent to which one's consumption of a good interferes with another's. A good is defined as non-rival if more consumption of it requires no additional cost (Cornes and Sandler 1986). People can simultaneously enjoy air and sunlight, watch television, and listen to radio. Adding consumers does not add cost; one's enjoyment does not interfere with another's. Few goods are perfectly non-rival. Many goods that are non-rival initially, such as road use, become less non-rival (and more rival) as congestion creeps in (Leach 2004, pp. 155-6).

 To early observers, software was a pure non-rival good. Once software was produced, the cost of sharing it appeared miniscule, because a contributor could keep the software while providing a perfect copy to someone else. Because one's benefit does not interfere with another's, "you never lose from letting your product free," exclaimed Ghosh (1998). Thus, non-rivalry became central in the discussion of open source software, an early instance of OC. Non-rivalry was frequently cited in discussing and modeling of the phenomenon (Baldwin et al. 2006; Harhoff et al. 2003, pp. 1759-67). Some scholars have argued that OC performs well because it produces "anti-rival goods" (Weber 2004), which benefit those who shares then.

But others have recognized that OC can involve goods that are somewhat rival, even if just because contributions can require attention, time, and effort (Shah 2006, p. 1005). If we consider that sharing may require an effort, than even software may not be the pure non-rival good it seems (Baldwin and Clark 2006; Marengo and Pasquali 2010). We aim to advance this discussion.

### 3.   RESULTS: OPEN COLLABORATION PERFORMS ROBUSTLY

To assess the performance of OC, we combine the three elements — cooperativeness, need heterogeneity, rivalry — in an agent-based model. Because we envision OC as a general system for innovation and production, we use a fundamental measure of economic performance: efficiency in turning inputs to outputs.

We find surprising results: OC can thrive even in seemingly harsh environments, reaching much more broadly than observers assumed or observed. It performs robustly even when cooperators are a fraction of participants, free riders are present, goods are rival, or participant needs are homogenous (non-diverse).The results are summarized in Figure 1, Figure 2, and Figure 3.

The results suggest what affects the performance of OC, and, equally important, what does not. Some conditions are commonly assumed necessary – but they are not. First, OC can thrive even if participants not an exclusive bunch of cooperators, but just a random sample from the human population, where cooperators are a small minority. Second, it is not required that participants derive immediate benefits from contribution, such as monetary gains, enhanced professional reputation, or pleasure. Third, OC is not derailed when the resources shared are rival or when participant needs are highly similar. It can perform well even with rival goods or homogeneous needs; performance suffers only when the two are concurrent. The model also explains some intriguing observations, such as the extreme disparity in contributions to OC, where a small core contributes the most and many contribute little.

The findings imply that OC is likely to grow and spread into new domains. They also inform the discussion on new organizational forms, collaborative and communal (e.g., Adler et al. 2008; Benkler 2011; Faraj et al. 2011; Heckscher and Adler 2006). Human efforts, it seems, can be harnessed in previously unthought-of ways: by relying on goal-oriented yet loosely coordinated participants, who interact to create products of economic value, which they then offer to anyone, contributors and free riders alike.

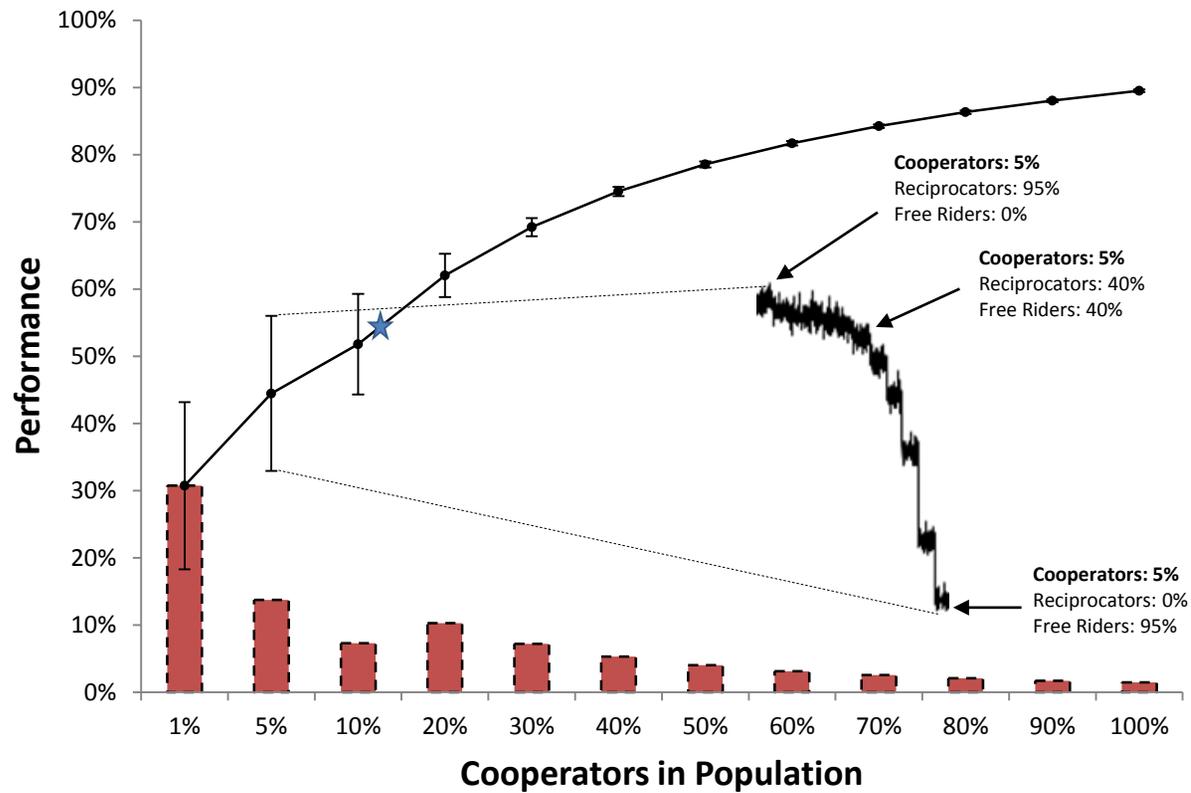

**Figure 1.** Mean performance by proportion of Cooperators in the population (solid line) with 95% confidence intervals. Bars show the marginal improvement in performance. Callout shows effect of Reciprocator and Free Riders composition when Cooperator proportion is fixed at 5%. The star signifies the ratio of cooperators in the general population (13%). Held constant were Rivalry (none) and Need Heterogeneity (high).

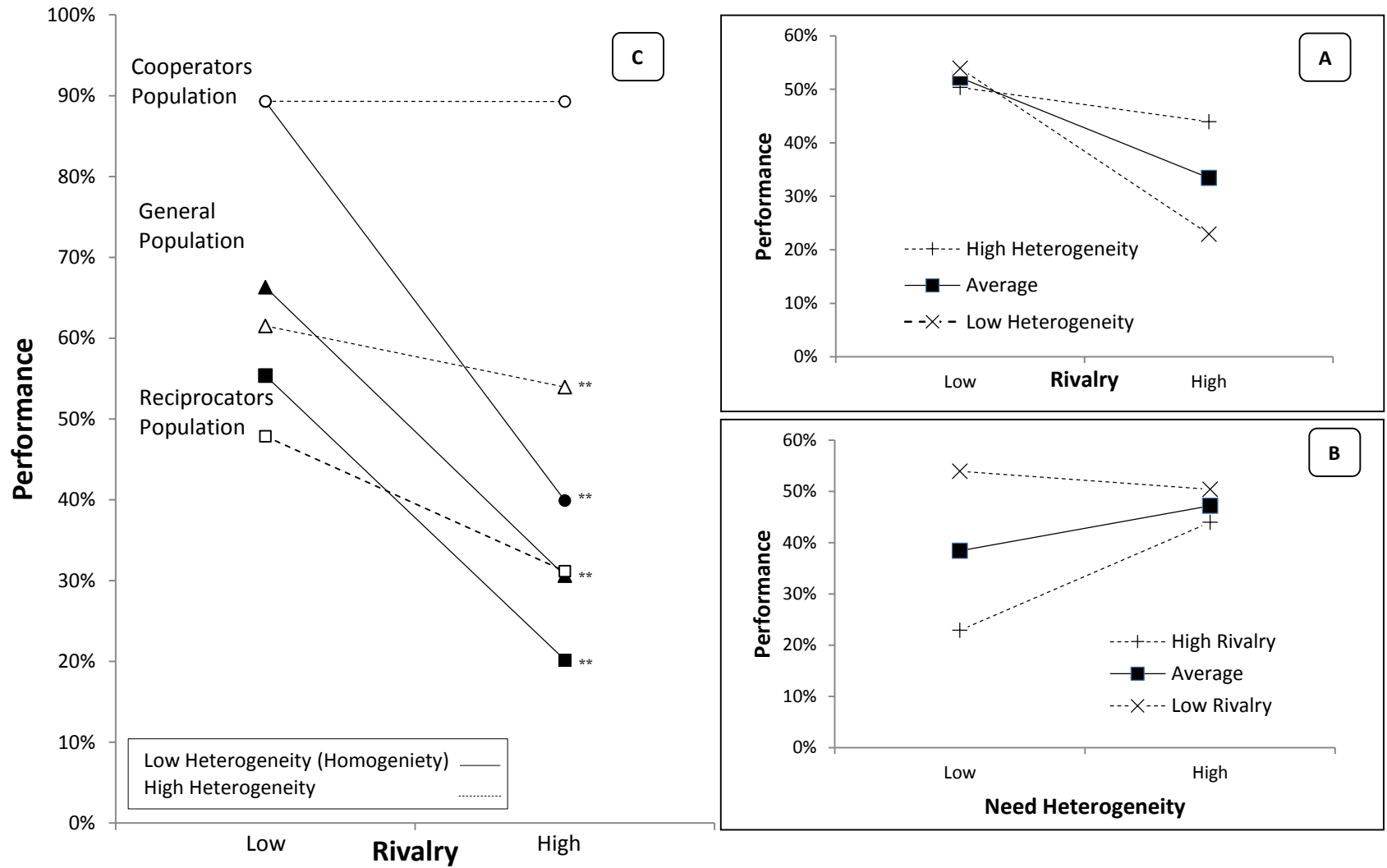

**Figure 2A, 2B, 2C. How Elements Interact to Affect Performance.** Main effect of Rivalry and interaction with Need Heterogeneity over a combination of three populations: cooperators, the general population and reciprocators **(A)**. Main effect of Need Heterogeneity and interaction with Rivalry over the same populations **(B)**. Interactions between Rivalry, Need Heterogeneity, and population composition **(C)**. Free-Rider populations (not plotted) showed lower performance (<10% in all conditions, p<0.001).

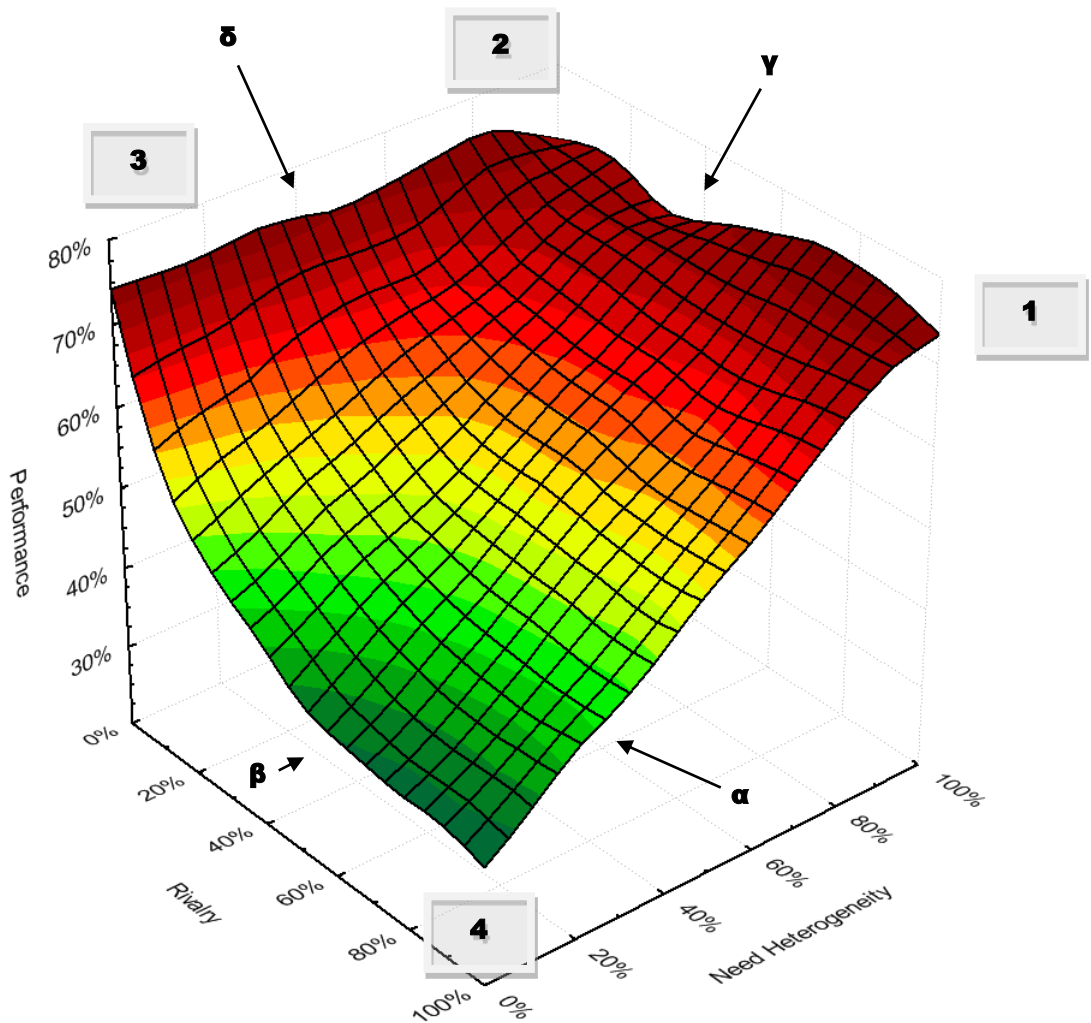

**Figure 3.** Performance contour plot for the general population at various combinations of Need Heterogeneity and Rivalry, ranging from high (red) to low (green). Post-hoc analysis (Tukey HSD) indicated that the four facets (corner points) differ significantly from each other (all p<0.001).